\documentclass[12pt]{article}
\usepackage[utf8]{inputenc}

\usepackage{authblk, titling}
\usepackage{graphicx}%
\usepackage{multirow}%
\usepackage{amsmath,amssymb,amsfonts,mathtools}%
\usepackage{mathrsfs}%
\usepackage[title]{appendix}%
\usepackage[dvipsnames]{xcolor}%
\usepackage{textcomp}%
\usepackage{manyfoot}%
\usepackage{booktabs}%
\usepackage{algorithm}%
\usepackage{algorithmicx}%
\usepackage{algpseudocode}%
\usepackage{listings}%
\usepackage{cite}
\usepackage{parskip}
\usepackage[left=2cm,right=2cm,top=3cm,bottom=3cm]{geometry}

\usepackage[linktocpage=true]{hyperref}

\usepackage{soul}

\usepackage{tikz}
\usetikzlibrary{arrows.meta}

\tikzset{three sided/.style={
        draw=none,
        append after command={
            [shorten <= -0.5\pgflinewidth]
            ([shift={(-1.5\pgflinewidth,-0.5\pgflinewidth)}]\tikzlastnode.north east)
        edge([shift={( 0.5\pgflinewidth,-0.5\pgflinewidth)}]\tikzlastnode.north west) 
            ([shift={( 0.5\pgflinewidth,-0.5\pgflinewidth)}]\tikzlastnode.north west)
        edge([shift={( 0.5\pgflinewidth,+0.5\pgflinewidth)}]\tikzlastnode.south west)            
            ([shift={( 0.5\pgflinewidth,+0.5\pgflinewidth)}]\tikzlastnode.south west)
        edge([shift={(-1.0\pgflinewidth,+0.5\pgflinewidth)}]\tikzlastnode.south east)
        }
    }
}

\usepackage{siunitx}
\usepackage[normalem]{ulem}

\newcommand{\A}{{\mathcal A}}
\newcommand{\B}{{\mathcal B}}
\newcommand{\C}{{\mathcal C}}
\def\pert{\varepsilon}
\def\manifold{\mathcal{M}}

\def\tidalpt{K_1}

\def\tidalG{G^{(1)}{}}
\def\tidalT{T^{(1)}{}}

\def\ith{{(i)}}
\def\ithp1{{(i+1)}}

\def\defor{\Xi}

\def\rstar{{R_N}}
\def\rbound{{R_i}}

\def\ro{R_*}

\def\diff{\mathrm{d}}

\def\pert{\varepsilon}
\def\manifold{\mathcal{M}}

\def\Gtt{G^t{}_t}

\def\Grr{G^r{}_r}

\def\Gthetatheta{G^\theta{}_\theta}

\def\Gphiphi{G^\phi{}_\phi}

\def\sGtt{[G^t{}_t]}

\def\sGrr{[G^r{}_r]}

\def\sGthetatheta{[G^\theta{}_\theta]}

\makeatletter
\newcommand*\bigcdot{\mathpalette\bigcdot@{.5}}
\newcommand*\bigcdot@[2]{\mathbin{\vcenter{\hbox{\scalebox{#2}{$\m@th#1\bullet$}}}}}
\makeatother

\title{Review on the matching conditions for the tidal problem:\\
towards the application to more general contexts}
\author{Eneko Aranguren\thanks{eneko.aranguren@ehu.eus}}
\author{Ra\"ul Vera\thanks{raul.vera@ehu.eus}}
\affil{Fisika Saila, University of the Basque Country UPV/EHU, \\
Barrio Sarriena s/n, 48940 Leioa, Basque Country}

\begin{document}
\maketitle

%%==========%%
%% Abstract %%
%%==========%%

\begin{abstract}
The tidal problem
is used to obtain the tidal deformability (or Love number)
of stars.
The semi-analytical study
is usually treated in perturbation
theory as a first order perturbation problem over a spherically symmetric
background configuration consisting of a stellar interior
region matched across a boundary to a vacuum exterior region that models the tidal field.
The field equations for the metric and matter perturbations
at the interior and exterior regions
are complemented with corresponding boundary conditions.
The data of the two problems at the common boundary are
related by the so called matching conditions.
These conditions for the tidal problem
are known in the contexts of perfect fluid stars
and superfluid stars modelled by a two-fluid.
Here we review the obtaining of the
matching conditions for the tidal problem
starting from a purely geometrical setting, and present
them so that they can be readily
applied to more general contexts,
such as other types of matter fields,
different multiple layers or phase transitions.
As a guide on how to use the matching conditions, we
recover the known results
for perfect fluid and superfluid neutron stars.
\end{abstract}

\tableofcontents

%\keywords{General Relativity, Perturbation Theory, Tidal Problem, Matching conditions}

%%%%=======================%%%%
%%%%       MAIN TEXT       %%%%
%%%%=======================%%%%

%%%%%%%%%%%%%%%%%%%%%%%%%%%%%%%%%%%%%%%%%%%%%%%%%
%%%%%%%%%%%%%%%% INTRODUCTION %%%%%%%%%%%%%%%%%%%
%%%%%%%%%%%%%%%%%%%%%%%%%%%%%%%%%%%%%%%%%%%%%%%%%
\section{Introduction}\label{sec:intro}

The purpose of this paper is to briefly review the
matching conditions in perturbation theory to first order involved in
the even-parity tidal problem, and present them in a way
that are readily applicable to generalizations other than in the known
perfect fluid
\cite{Damour:2009vw,Hinderer:2010,reina-sanchis-vera-font2017}
or two-fluid (to model superfluid) stars \cite{Aranguren:tidal} contexts. 

The aim of studying the tidal problem 
is to obtain the Love number, or, equivalently,
the tidal deformability, of stars.
The global problem consists of a region to model the stellar
interior which is matched to an exterior vacuum region
that models a tidal field,
produced e.g. by a companion star.
Using perturbation theory the model is built
as a first order perturbation on top
of a static and spherically symmetric background configuration
consisting of an interior region ball of radius $\ro$ with
a Schwarzschild exterior.
The Love number is then determined from the value
that certain function
of the first order metric perturbation,
that depends only on the radial coordinate, $y^-(r)$,
takes at the outer surface (the surface as seen
from the exterior) of the star, specifically $y^-(\ro)$.
That value is found by integrating an analogous
function $y^+(r)$ in the interior region from the origin
outwards to obtain $y^+(\ro)$,
and then use whatever information
we have on the jump $[y]:=y^+(\ro)-y^-(\ro)$
to determine $y^-(\ro)$ (see \cite{Hinderer:2008}).

In the context of perfect fluid models, it was argued in
\cite{Damour:2009vw, Hinderer:2010}
that the jump
is given by (see Eq.~(15) in \cite{Hinderer:2010})
\begin{equation}\label{eq:matching_y}
    [y]=\varkappa\frac{\ro^3}{2M} E(\ro),
\end{equation}
where $E$ is the energy density (of the background configuration),
$M$ is the mass of the star, and we use $\varkappa$
for the gravitational coupling constant ($\varkappa=8\pi$ in
natural units $G=c=1$).
The argument uses the fact that the function $y(r)$ is
defined as the quotient $r H'(r)/H(r)$, where $H$ is a function
that describes part of the first order perturbation
(that is the harmonic $\ell=2$ part of the $H_0$ function
of the Regge-Wheeler decomposition \cite{regge:1957}). The problem
is set on the whole domain $r\in(0,\infty)$, and $H$ is implicitly assumed to be continuous at $r=\ro$,
just as well as the rest of the perturbation metric functions.
The key idea is that the Einstein field equations (EFE) at first order
can be defined (in strict terms at least in a distributional sense),
and $H(r)$ thus satisfies a second order ODE that
contains a term with $dE/dP$, where
$P$ is the
pressure
of the background configuration.
Therefore, since $P(r)$ must be continuous at the surface,
if $E(r)$ has a jump at $r=\ro$, then $dE/dP$ presents a Dirac delta term
there proportional to $E(\ro)$. 
As a result, the solution of $H(r)$ yields a continuous function
but discontinuous $H'(r)$ at $r=\ro$, with a jump $[H']$
proportional to $E(\ro)$. Completing the chain
$[y]=[r H'/H] = \ro [H'] / H(\ro)$ with the explicit
expression of $[H']$ leads to \eqref{eq:matching_y}.

Although the result is correct, the procedure has two drawbacks.
The first comes from a practical point of view. The procedure relies
completely on the use of the EFEs for a perfect fluid at both sides
(vacuum taken as a trivial particular case).
If one needs to consider other matter fields at the interior,
or even phase transitions in different regions, the derivation of $[y]$
must be carried out for each different case.
Let us stress that after \cite{Damour:2009vw,Hinderer:2010}
were published,
several works on other types of stellar interiors miss
the derivation of $[y]$
to properly justify that
$y$ is taken to be ``continuous".

The second drawback is conceptual, since the procedure also relies
on the setting of the interior and exterior problems
as one single problem on a common domain, $r\in(0,\infty)$, and then implicitly
assumes that the perturbed metric functions have the continuity
properties needed to devise the EFEs in a distributional sense.
This setting has its basis in the original Hartle-Thorne perturbative framework \cite{Hartle1967,Hartle2}, where all the perturbed metric functions
were assumed to be continuous in that sense.
However, the study of matchings in perturbation theory has shown that
that construction is not necessary, and not desirable in some
cases of interest.
In perturbed matching theory,
the interior and exterior first order problems
(denoted with a $+$ and $-$ respectively)
are more conveniently
treated as two separate (gauge field) problems
with related boundary data at
the common boundary $r_+=r_-=:\ro$,
with basis on a purely geometrical construction.
In fact, some of the functions may indeed present jumps,
that can be made to vanish by partially fixing the gauges,
whereas some others necessarily present jumps in general,
as otherwise the setting becomes inconsistent.
In particular, the amendment
needed to the original Hartle-Thorne model
has consequences in the computation of the stellar mass to second order \cite{ReinaVera2015}.

Although perturbed matching theory may seem to be a mere mathematical artifact,
apart from providing firm and rigorous grounds to the results on perturbed matchings
based on the Hartle-Thorne framework (after the needed amendments),
it
provides, on the one hand, full control
over the gauges at either side independently
(see \cite{Marc_Filipe_Raul_2007}).
This fact is key in the
proofs of uniqueness and existence of compact
rotating configurations in GR in perturbation theory to second order
\cite{MRV1,MRV2}.
On the other hand, the geometrical basis provides a direct way to generalize the
derivation, in particular, of $[y]$ to any matter field content,
even including different kind of layers and phase transitions.

In this paper we review the
matching conditions to first order in perturbation theory
(based on the works \cite{Mars:2005,ReinaVera2015,MRV2})
aimed at the tidal problem, starting from a pure geometrical setting,
and thus ready
to be used to obtain the matching conditions for general stellar
matter field contents.
We take the opportunity to include a formal way
of dealing with multiple concentrically distributed regions.
To show how to use the conditions,
we review the obtaining of the perturbed matching conditions
to first order for the tidal problem
for
perfect fluid stars \cite{reina-sanchis-vera-font2017},
thus recovering (and putting on firm grounds) the condition
\eqref{eq:matching_y}, and two-fluid
superfluid stars \cite{Aranguren:tidal}.
For completeness, we also provide in Sect. \ref{sec:love_number}
the usual procedure to compute the Love number
and how the jump $[y]$ enters the calculation.

We include in the following subsection a brief account on
the matching in perturbation theory to
set the grounds, provide some relevant references, and fix some notation.

\subsection{Matching in perturbation theory}
In essence, the problem of matching in perturbation theory
to first order in General Relativity starts with a set of two spacetimes with
boundary $(\manifold^+,g^+,\Sigma^+)$ and
$(\manifold^-,g^-,\Sigma^-)$ that have been matched across
$\Sigma^+=\Sigma^-=:\Sigma$ to create a single spacetime
$(\manifold,g)$ with two regions, where
$\manifold=\manifold^+\cup\manifold^-$
and $g$ equals $g^+$ or $g^-$ on each corresponding region.
Each region is now endowed with a symmetric tensor,
$\tidalpt^+$ and $\tidalpt^-$,
which describe the metric perturbation on the
corresponding region.
The perturbation parameter $\pert$ is chosen
so that the families of metrics
$g_\pert^\pm:=g^\pm+\pert\tidalpt^\pm$
describe the metric to first order at each region.
By taking the metric $g_\pert$ on $\manifold$ defined to be
$g_\pert^+$ on $\manifold^+$ and $g_\pert^-$ on $\manifold^-$,
one can proceed to construct its Riemann tensor
$\mathrm{Riem}(g_\pert)$. We say that the two
regions of the given background $(\manifold,g)$
perturbed with $\tidalpt^\pm$
match (to first order)
when $\mathrm{Riem}(g_\pert)$ to first order in $\pert$
can be constructed as a
distribution \emph{and} does not have a Dirac delta term.
The general treatment of the matching to first order,
explicitly applicable in terms of any gauge,
was done in \cite{Battye01} and \cite{Mukohyama00}
 (see e.g. \cite{JMaria-Carsten} and \cite{Marc_Filipe_Raul_2007}
 regarding treatments in terms of gauge invariants).
It was shown
that the two perturbed regions match to first order
if and only if there exist
a couple vectors $Z_1^\pm$ defined at points on $\Sigma$ so that
a certain set of conditions on $\Sigma$ are satisfied, which depend on
$\{g^\pm,\tidalpt^\pm,Z_1^\pm\}$. We refer to those as the
first order perturbed matching conditions,
and the normal part of $Z_1^\pm$ to $\Sigma$, that we shall denote
by $\Xi_1^\pm$, describe the  deformation of $\Sigma$
(to first order)
as seen from each side \cite{Mars:2005} in terms
of the gauges (or class of gauges) chosen.

The second order problem is constructed analogously in terms
of an extra symmetric tensor for each region $K_2^\pm$
to describe the second order perturbations, and the
corresponding second order matching conditions (found
in \cite{Mars:2005}) demand the existence of two
extra vectors $Z_2^\pm$ such that
a certain set of conditions for
$\{g^\pm,\tidalpt^\pm,K_2^\pm,Z_1^\pm,Z_2^\pm\}$
on $\Sigma$ are satisfied.

The particularization
to second order stationary and axially symmetric perturbations around
static and spherically symmetric backgrounds was performed
in \cite{ReinaVera2015} (see also \cite{MRV2}). It must be stressed
that the
perturbation
tensor that suits the even-parity tidal problem
scenario enters those constructions at second order.
The first order matching conditions suitable for
the tidal problem correspond
to
the conditions at second order
with a vanishing first order problem
in the treatment in
\cite{ReinaVera2015,MRV2}
(see \cite{reina-sanchis-vera-font2017,Aranguren:tidal}).

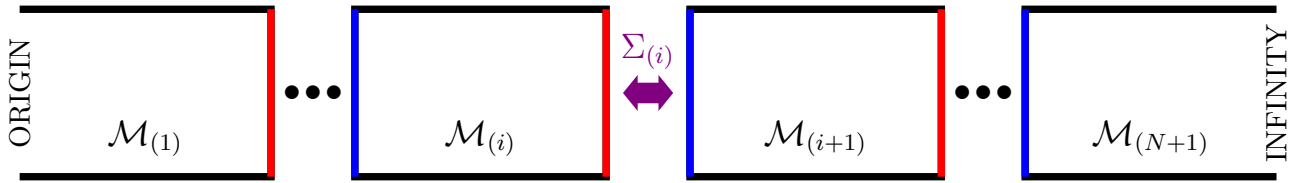
\begin{figure*}
\resizebox{\textwidth}{!}{
\begin{tikzpicture}\label{fig:multiple_boundaries}

%% FIRST BOX
\draw[black, line width=2.5pt] (0,0) -- (3,0) -- (3,2) -- (0,2);
\draw[red, line width = 2.5pt] (3,0) -- (3,2);

\node[rotate=90] at (0,1) {\footnotesize{ORIGIN}};
\node[above] at (1.5,0.1) {$\mathcal{M}_{(1)}$};

\node at (3.25,1) {$\bullet$};
\node at (3.5,1) {$\bullet$};
\node at (3.75,1) {$\bullet$};

%% SECOND BOX
\draw[black, line width=2.5pt] (4,0) rectangle (7,2);
\draw[blue, line width = 2.5pt] (4,0) -- (4,2);
\draw[red, line width = 2.5pt] (7,0) -- (7,2);

\node[above] at (5.5,0.1) {$\mathcal{M}_{\ith}$};
\node at (7.5,1.5) {$\color{violet}{\Sigma_{(i)}}$};

\draw[{Triangle[width=12pt,length=5pt]}-{Triangle[width=12pt,length=5pt]}, line width=7pt, color=violet] (7.2, 1) -- (7.8,1);

%% THIRD BOX
\draw[black, line width=2.5pt] (8,0) rectangle (11,2);
\draw[blue, line width = 2.5pt] (8,0) -- (8,2);
\draw[red, line width = 2.5pt] (11,0) -- (11,2);

\node[above] at (9.5,0.1) {$\mathcal{M}_{\ithp1}$};

\node at (11.25,1) {$\bullet$};
\node at (11.5,1) {$\bullet$};
\node at (11.75,1) {$\bullet$};

%% FOURTH BOX
\draw[black, line width=2.5pt] (15,0) -- (12,0) -- (12,2) -- (15,2);
\draw[blue, line width = 2.5pt] (12,0) -- (12,2);

\node[above] at (13.5,0.1) {$\mathcal{M}_{(N+1)}$};
\node[rotate=90] at (15,1) {\footnotesize{INFINITY}};

\end{tikzpicture}
}
\caption{Schematic representation of the matching between different manifolds. The $i$-th manifold $\manifold_{\ith}$ is matched to $\manifold_{\ithp1}$ through the hypersurface $\Sigma_{\ith}$.
At $\Sigma_{\ith}$
$\manifold_{\ith}$ plays the role of the inner part of the matching,
$\manifold^+=\manifold_{(i)}$, and $\manifold_{(i+1)}$ the
outer part, so that $\manifold^-=\manifold_{(i+1)}$.
The innermost and outermost regions, $\manifold_{(1)}$ and $\manifold_{(N+1)}$ are not linked to any other spacetime.}
\end{figure*}

The local nature of the matching procedure allows
us to trivially devise the (perturbed) matching procedure needed for a stellar interior
made up of  layers with different matter contents.
For the sake of formality
we consider
a construction based on
a set of $N+1$
regions $\{\manifold_{\ith}\}$
with boundary
where $i=1,...,N+1$, so that $\manifold_{(1)}$ contains the origin
and are ordered from inner to outer, matched together
across the matching hypersurfaces
$\Sigma_{\ith}$ with $i=1,\ldots,N$
to form a global static and spherically
symmetric background spacetime as depicted in Fig.~\ref{fig:multiple_boundaries}.
Note that one may include $\Sigma_{(N+1)}$ to denote ``infinity''.
At any $\Sigma_{\ith}$ separating two regions, say $\manifold^+=\manifold_{_{\ith}}$ (inner)
and $\manifold^-=\manifold_{\ithp1}$ (outer), for any pair of functions $f^+$ and $f^-$
defined on the corresponding region,
we will use $[f]_i:=f^+\lvert_{\Sigma_{\ith}}-f^-\lvert_{\Sigma_{\ith}}$
to denote the ``jumps'' of $f$ there.

%%%%%%%%%%%%%%%%%%%%%%%%%%%%%%%%%%%%%%%%%%%%%%%%%
%%%%%%%%%%%% GEOMETRICAL MATCHING %%%%%%%%%%%%%%%
%%%%%%%%%%%%%%%%%%%%%%%%%%%%%%%%%%%%%%%%%%%%%%%%%
\section{Geometrical perturbed matching conditions}\label{sec:geometrical}
In this section we set forth the perturbed matching conditions
to first order for
static and axially symmetric
even parity perturbations
around a static and spherically symmetric background
at any hypersurface
$\Sigma_{\ith}$ in purely geometrical terms.
With the tidal problem in mind we restrict the analysis, as usual,
  to the $\ell\geq 2$ sector.
We refer to \cite{Aranguren:tidal}, based in turn on \cite{MRV2} and \cite{ReinaVera2015},
for the proving details.

In principle, each $\manifold_{\ith}$ is endowed with the usual
spherical coordinates $\{t_i,r_i,\theta_i,\phi_i\}$,
and we may indicate by $f_i$ that a function is defined on $\manifold_{\ith}$.
However, to avoid flooding all equations with $i$ indices, in what follows
we only show explicitly the $i$-dependence at
the jumps of
functions.

%%%%%%%%%%%% BACKGROUND %%%%%%%%%%%%%%%
\subsection{Background}
The metric of the background configuration in spherical coordinates is given by
\begin{align}
  g=-e^{\nu(r)}\diff t^2+e^{\lambda(r)}\diff r^2+r^2(\diff\theta^2+\sin^2\theta \diff\phi^2),\label{eq:metric}
\end{align}
($i$-subindexes ommited in the functions and coordinates)
on each $\manifold_{(i)}$.
Given the spherical symmetry of the
whole configuration, the matching hypersurfaces
$\Sigma_{\ith}$ inherit the symmetries
and they can be parametrized by common values
of $\{t,\theta,\phi\}$ at each side (see e.g. \cite{Vera2002}).
Then, the matching conditions for the background configuration are 
    $[r]_i=0$,
that establishes that $\Sigma_{(i)}$ is defined by the same value, say $R_i$,
of the respective radial coordinates $r$, and
\begin{align}
  &[\lambda]_i=0,\quad [\nu]_i=0,\label{eq:s_background}\\
  &[\nu']_i=0.\label{eq:s_dnu_geo}
\end{align}
Observe that the condition $[\nu]_i=0$ is, in fact, a consequence
of the choice of the $t$ coordinate at each side, which have been
taken so that their values coincide on $\Sigma_{\ith}$.
It is in that sense that we say that the matching condition
just serves to accommodate the choice of ``gauges'' on each side.

%%%%%%%%%%%% TIDAL %%%%%%%%%%%%%%%
\subsection{First order}
We take the first order perturbation tensor
$\tidalpt$ describing the static, axially symmetric and even parity tidal field as given
in the Regge-Wheeler gauge, and decomposed in Legendre polynomials,
\begin{align}
  \tidalpt = \sum_{\ell}\left\{e^{\nu(r)}H_{0 \ell}(r)\diff t^2 + e^{\lambda(r)}H_{2 \ell}(r)\diff r^2+ r^2 K_{\ell}(r)(\diff\theta^2 + \sin^2\theta\diff\phi^2)\right\}P_{\ell}(\cos \theta),\label{tidal:tidalpt}
\end{align}
on each region.
The matching conditions for $\ell\geq2$ on each $\Sigma_{(i)}$
read [see Eqs.~(37)-(39) in \cite{Aranguren:tidal}]
\begin{align}
    &[K_{\ell}]_i=0,\quad [H_{0\ell}]_i=0, \label{tidal:s_K_H0_1} \\
    &[H_{2\ell}]_i-\rbound [K_{\ell}']_i= e^{-\lambda(\rbound)/2}\defor_{\ell i}[\lambda']_i,\label{tidal:s_H2_dK_1}\\
    &[H'_{0\ell}]_i+\frac{R_i}{2} \nu'(\rbound)[K'_\ell]_i=
      -e^{-\lambda(\rbound)/2}\defor_{\ell i}[\nu'']_i,\label{tidal:s_dH0_dK_1}
\end{align}
for some functions $\defor_{\ell i}$ at each $\ell\geq 2$.
As mentioned in the Introduction, the functions $\defor_{\ell i}$
will describe
the deformation of the hypersurface $\Sigma_{(i)}$ in the class of gauges
compatible with the problem
(and it is equal at both sides) \cite{ReinaVera2015, MRV2}.
The first order problems for $\ell\geq 2$ will then match at $\Sigma_{(i)}$
if and only if there exists a quantity $\Xi_{\ell i}$ such that
Eqs.~\eqref{tidal:s_K_H0_1}-\eqref{tidal:s_dH0_dK_1} are satisfied.

Although the matching conditions for $\ell=0,1$ are irrelevant as far as the tidal
problem is concerned, they can be obtained from \cite{ReinaVera2015} or \cite{MRV2}, and find, in particular, that the jumps become arbitrary enough as to accommodate all the gauge freedom left at this point.

%%%%%%%%%%%%%%%%%%%%%%%%%%%%%%%%%%%%%%%
%%%%%%%%% MATCHING FROM G %%%%%%%%%%%%%
%%%%%%%%%%%%%%%%%%%%%%%%%%%%%%%%%%%%%%%
\subsection{Matching conditions in terms of the Einstein tensor}
Our goal now is to express the matching conditions in terms of the Einstein tensor
of the background geometry $G^\alpha{}_\beta:= \mathrm{Ein}(g)^\alpha{}_\beta$.
The only independent and nonzero components satisfy the relations
(we drop the $r$-dependence here, prime denotes derivative with respect to $r$)
\begin{align}
    &\lambda'=+\frac{1-e^\lambda}{r}-r e^\lambda \Gtt,\label{eq:dlambda}\\
    &\nu'=-\frac{1-e^\lambda}{r}+r e^\lambda \Grr,\label{eq:dnu}\\
    &\nu'' = \frac{1}{2r}(\lambda'-\nu')(2+r\nu')+2 e^{\lambda} \Gthetatheta, \label{eq:ddnu}
\end{align}
on any region $\manifold_{\ith}$. Computing the jumps
of the above relations on any $\Sigma_{\ith}$
yields
\begin{alignat}{2}
    &[\lambda']_i=-\rbound e^{\lambda(\rbound)}\sGtt_i,\label{eq:s_dlambda}\\
    &[\nu']_i=+\rbound e^{\lambda(\rbound)}\sGrr_i,\label{eq:s_dnu}\\
    &[\nu'']_i=\left(1+\frac{\rbound\nu'(\rbound)}{2}\right)\frac{[\lambda']_i}{\rbound}+2 e^{\lambda(\rbound)}\sGthetatheta_i.\label{eq:s_ddnu}
\end{alignat}
As a result, using also that the mass function is given by
\begin{align}
    M(r)=\frac{r}{2}\left(1-e^{-\lambda(r)}\right),\label{eq:mass}
\end{align}
the matching conditions
\eqref{eq:s_background}-\eqref{eq:s_dnu_geo}
can be rewritten as
\begin{align}
  &[M]_i=0,\quad[\nu]_i=0,
  \label{eq:matching_background1}\\
  &[\Grr]_i=0, \label{eq:matching_background2}
\end{align}
while
\eqref{tidal:s_K_H0_1}-\eqref{tidal:s_dH0_dK_1}
read
\begin{align}
    &[K_{\ell}]_i=0,\quad [H_{0\ell}]_i=0,\label{tidal:s_K_H0_2}\\
    &[H_{2\ell}]_i-\rbound [K_{\ell}']_i= -\rbound e^{\lambda(\rbound)/2}\defor_{\ell i}\sGtt_i,\label{tidal:s_H2_dK_2}\\
    &[H'_{0\ell}]_i-[K'_\ell]_i= 
    -2 e^{\lambda(\rbound)/2} \defor_{\ell i}\sGthetatheta_i-\left(1+\frac{\rbound\nu'(\rbound)}{2}\right)\frac{[H_{2\ell}]_i}{\rbound}.\label{tidal:s_dH0_dK_2}
\end{align}

%%%%%%%%%%%%%%%%%%%%%%%%%%%%%%%%%%%%
%%%%%%% GENERAL RELATIVITY %%%%%%%%%
%%%%%%%%%%%%%%%%%%%%%%%%%%%%%%%%%%%%
\section{Application to General Relativity}

In this section we review how the above construction
of the perturbed matching to first order
applies to problems in perturbation theory to first order in
General Relativity (without cosmological constant)
in the present context for the tidal problem,
i.e. perturbations of the form \eqref{tidal:tidalpt}
around static and spherically symmetric backgrounds. 
For any general case, anyway, the procedure consists in the three
following steps.
\begin{enumerate}
\item[(i)] As a first step, since only the Einstein tensor \emph{of the background geometry}
enters the perturbed matching to first order,
it suffices to introduce
the  Einstein field equations (EFE) at the background level (we omit the $i$ index
for each region)
\begin{align}
    G{}^\alpha{}_\beta = \varkappa T{}^\alpha{}_\beta,\label{eq:efes_back}
\end{align}
in the equations
\eqref{eq:matching_background1}-\eqref{tidal:s_dH0_dK_2}
to obtain the geometrical perturbed matching conditions in terms of
quantities related to the desired matter field contents of the model.
We are using $G=\mathrm{Ein}(g)$, $T$ for the energy-momentum tensor
at the background level, and $\varkappa$ for the gravitational coupling constant.

\item[(ii)] The second step consists in using the EFEs at first order, together
with the perturbed matching conditions from the first step.
This may provide additional conditions on the jumps of the
perturbation metric functions $\{H_{0\ell},H_{2\ell}, K_{\ell}\}$.

\item[(iii)] The final step corresponds to the addition of the
  matter field matching conditions
  governing
  the behaviour of the matter fields across layers, such as
  surfaces separating different media, phase transitions,
  charged surfaces, etc...
\end{enumerate}

Before we start with the first step,
observe that
the Einstein tensor of the background geometry given by \eqref{eq:metric}
only has (at most) three non-vanishing components
$G^t{}_t$, $G^r{}_r$, and $G^\theta{}_\theta=G^\phi{}_\phi$ (in all regions).
The form of the energy-momentum tensor compatible with that, c.f. \eqref{eq:efes_back},
is many times referred to as an ``anisotropic fluid'' with ``radial pressure''
$T^r{}_r=\varkappa^{-1}G^r{}_r$. Equation \eqref{eq:matching_background2}
thus states that the radial pressure cannot have a jump on any matching
hypersurface $\Sigma_{\ith}$.
If the outermost region is the exterior vacuum to a compact object, this condition
establishes the value $\rstar$ of the radial coordinate where the boundary
common to the two problems is located.
On the other hand, notice, however, that neither
$[G^\theta{}_\theta]_{i}$ nor  $[G^t{}_t]_{i}$
are necessarily zero in general, and therefore neither
are $[T^\theta{}_\theta]_{i}$ nor $[T^t{}_t]_{i}$.
If there are no equations for the matter fields,
nor additional field matching conditions
for them,
the necessary and sufficient conditions for ``anisotropic fluids''
to match at first order are the background matching equations
\eqref{eq:matching_background1}-\eqref{eq:matching_background2}
plus the two first order conditions in \eqref{tidal:s_K_H0_2}. Therefore, in that case,
the jumps of $H_{2\ell}$ and the derivatives
of all the perturbation metric functions are not constrained, a priori.

In the following we consider perfect fluid
and two-fluid superfluid regions,
using steps (i) and (ii) to
recover the known matching conditions
in the tidal problem for neutron and quark stars
(see e.g. \cite{Damour:2009vw,Hinderer:2010,reina-sanchis-vera-font2017})
and superfluid neutron stars (see e.g. \cite{Aranguren:2022}).

\subsection{Step (i): geometrical perturbed matching conditions
  using the background field equations}

Let us start by performing the first step (i)
for perfect fluid and two-fluid regions.

\noindent
\emph{Perfect fluid:} To fix some notation, we write
the energy-momentum tensor for a perfect fluid as
\begin{align}
    T^\alpha{}_{\beta}=(E+P)U^\alpha U_\beta+P \delta^\alpha_{\beta},\label{eq:energy_momentum_background_PF}
\end{align}
for some unit fluid flow $U$,
where $E$ and $P$ are the corresponding energy density and pressure, respectively.
Given the form of the Einstein tensor
for \eqref{eq:metric}, as mentioned above,
the EFEs on the background imply that $U=e^{-\nu/2}\partial_t$,
and require only the equation
\begin{equation}
  G^r{}_r=G^\theta{}_\theta.\label{eq:back_pf}
\end{equation}
The energy density and pressure are then given by the relations
\begin{align}
    \Gtt=-\varkappa E,\quad\Grr(=\Gthetatheta=\Gphiphi)=\varkappa P.
\end{align}
The matching conditions \eqref{eq:matching_background1}-\eqref{tidal:s_dH0_dK_2}
applied to a perfect fluid thus read
\begin{align}
  &[M]_i=0,\quad[\nu]_i=0,\label{eq:matching_1}\\
  &[P]_i=0,\label{eq:matching_pressure}\\
    &[K_{\ell}]_i=0,\quad [H_{0\ell}]_i=0,\label{eq:matching_2}\\
    &[H_{2\ell}]_i-R_i [K_{\ell}']_i= \varkappa R_i e^{\lambda(R_i)/2}\defor_{\ell i}[E]_i,\label{eq:matching_PF_defor}\\
    &[H'_{0\ell}]_i-[K'_\ell]_i= -\left(1+\frac{R_i\nu'(R_i)}{2}\right)\frac{[H_{2\ell}]_i}{R_i}.\label{eq:matching_3}
\end{align}
Observe that because of the field equations at the background level,
c.f. \eqref{eq:back_pf},
we now have $[G^\theta{}_\theta]_i=0$ and therefore \eqref{tidal:s_dH0_dK_2},
that becomes \eqref{eq:matching_3}, involves perturbation metric functions
only.

If the outer region is vacuum, the jump of the energy is simply
$[E]_{i=N}=E^+(\rstar)$, where $\rstar$ satisfies
$P^+(\rstar)=0$.
In general, the values of the energy density at the matching hypersurfaces
$\Sigma_{\ith}$, and thus the jumps $[E]_i$, will be determined
by the barotropic EOS $E=E(P)$ that govern the stellar configurations
on the different regions.
In other cases with no barotropic EOS, such as
homogeneous stars \cite{Reina:2015jia} and some
Skyrme stars \cite{alberto_skyrme}, those values will be
determined by whatever relations used to close the system of equations.

Note that at this point the function $H_{2\ell}$ may present jumps
at the matching hypersurfaces (see \eqref{eq:matching_3}).
Later we are going to see how the perfect fluid equations
at first order (in combination with the rest of conditions)
imply the vanishing of those jumps.\\

\noindent
\emph{Superfluid:}
For completeness,
let us introduce very briefly the two-fluid superfluid formalism,
as described in \cite{Langlois:1998} (see also \cite{Comer1999}). For brevity, we refer to this formalism simply as \textit{superfluid}.
The flow of neutrons and protons is given by $n^\alpha=n u^\alpha$
and $p^\alpha=p v^\alpha$, where $n$ and $p$ are the number
densities of neutrons and protons, respectively, and $u^\alpha$ and $v^\alpha$ are
unit timelike vectors.
All the model is determined by the master function
\[
    \Lambda = \Lambda(n^2,p^2,x^2),
\]
where $n^2:=n_\alpha n^\alpha$, $p^2:=p_\alpha p^\alpha$,
and $x^2:=-p_\alpha n^\alpha$ is the interaction term.
With the help of the definitions
\begin{equation*}
    \mu_{\alpha}:=\B n_{\alpha}+\A p_{\alpha},\qquad\chi_{\alpha}:=\C p_{\alpha}+\A n_{\alpha},
  \end{equation*}
with
\begin{equation*}
    \A:=-\frac{\partial\Lambda(n^2,p^2,x^2)}{\partial x^2},\quad
    \B:=-2\frac{\partial\Lambda(n^2,p^2,x^2)}{\partial n^2},\quad
    \C:=-2\frac{\partial\Lambda(n^2,p^2,x^2)}{\partial p^2},
\end{equation*}
and
\begin{equation}\label{eq:generalizedpressure}
    \Psi:=\Lambda-n^{\alpha}\mu_{\alpha}-p^{\alpha}\chi_{\alpha},
\end{equation}
the energy-momentum tensor reads
\begin{equation}
  T^\alpha{}_\beta=\Psi \delta^\alpha_\beta+p^\alpha \chi_\beta+n^\alpha \mu_\beta.\label{eq:energymom_SF}
\end{equation}
Similarly as in the perfect fluid case, staticity and spherical symmetry
of the background geometry requires that $u^\alpha=v^\alpha$.
That induces \eqref{eq:energymom_SF} to take the form of a perfect fluid
with $E$ replaced by $-\Lambda$. As a result, 
for this two-fluid model the perturbed matching conditions
are given by the
set $\{\eqref{eq:matching_1},\eqref{eq:matching_2},\eqref{eq:matching_3}\}$
plus
\begin{align}
  &[\Psi]_i=0,\label{eq:matching_SF_psi}\\
    &[H_{2\ell}]_i-R_i [K_{\ell}']_i= -\varkappa R_i e^{\lambda(R_i)/2}\defor_{\ell i}[\Lambda]_i.\label{eq:matching_SF_defor}
\end{align}
Now, if the outer region is vacuum, then we have $[\Lambda]_{i=N}=\Lambda^+(\rstar)$,
where $\rstar$ satisfies $\Psi^+(\rstar)=0$.\\

\noindent
\emph{Superfluid and perfect fluid:}
Some models consider stellar configurations composed of different kinds of fluids.
For instance, in \cite{datta_char:2020} a stellar model with a superfluid core
and perfect fluid crust is studied. In such case,
the same arguments as those above for each side can be used to show that
the perturbed matching conditions for an inner superfluid region
and an outer perfect fluid region are given by the
set $\{\eqref{eq:matching_1},\eqref{eq:matching_2},\eqref{eq:matching_3}\}$
plus
\begin{align}
    &\Psi(R_i)=P(R_i),\label{eq:matching_SF_PF_1}\\
    &[H_{2\ell}]_i-R_i [K_{\ell}']_i=-\varkappa R_i e^{\lambda(R_i)/2}\defor_{\ell i}\left(\Lambda(R_i)+E(R_i)\right).\label{eq:matching_SF_PF_2}
\end{align}
If the star were composed of an inner perfect fluid and an outer superfluid
(even though this might not be physically realistic),
Eq.~\eqref{eq:matching_SF_PF_2} ought to be replaced by the same equation
with a change of sign at the right-hand side.

%%%%%%%%%%%%%%%%%%%%%%%%%%%%%%%%%%%%
%%%%%%%%% FIELD EQUATIONS %%%%%%%%%%
%%%%%%%%%%%%%%%%%%%%%%%%%%%%%%%%%%%%
\subsection{Step (ii): adding the field equations at first order}

Let us recall how the EFEs at first order
are derived from $G_\pert:=\mbox{Ein}(g_\pert)$
and some $\pert$-family of energy-momentum tensors $T_{\pert}$
such that $T_{\pert=0}=T$.
It suffices to consider
the family of equations
$
    G_\pert{}^\alpha{}_\beta = \varkappa T_\pert{}^\alpha{}_\beta,
$
that reduces to \eqref{eq:efes_back} for $\pert=0$.

The first order equations are thus given by
\begin{equation}
G^{(1)}{}^\alpha{}_\beta = \varkappa T^{(1)}{}^\alpha{}_\beta,\label{eq:efes_first}
\end{equation}
where
\begin{align*}
    \tidalG^\alpha{}_\beta=\frac{\partial G_\pert{}^\alpha{}_\beta}{\partial\pert}\bigg\lvert_{\pert=0},\quad \tidalT^\alpha{}_\beta = \frac{\partial T_\pert{}^\alpha{}_\beta}{\partial \pert}\bigg\lvert_{\pert=0}.
\end{align*}

For the perfect fluid, one takes the form
$
    T_\pert^\alpha{}_\beta = (E_\pert+P_\pert)U_\pert^\alpha U_{\pert \beta}+P_\pert \delta^\alpha_\beta,
$
for some vector $U_\pert$ unit with respect to $g_\pert$ and such that
$U_{\pert=0}=U$, and corresponding energy density $E_\pert$ and pressure $P_\pert$
satisfying $E_{\pert=0}=E$, $P_{\pert=0}=P$.
If a barotropic EOS  is given in the form $E(P)$,
then it is imposed that $E_\pert=E(P_\pert)$. 
Similarly, for the two fluid one considers
$n_\pert{}^\alpha=n_\pert u_\pert{}^\alpha$, $p_\pert{}^\alpha=p_\pert v_\pert{}^\alpha$
and, given a ruling master function $\Lambda(n^2,p^2,x^2)$,
takes $\Lambda_\pert=\Lambda(n_\pert^2,p_\pert^2,x_\pert^2)$
from where the rest of quantities are constructed, following the formalism
explained above,
to form
$
T_\pert{}^\alpha{}_\beta=\Psi_\pert \delta^\alpha_\beta+p_\pert{}^\alpha \chi_\pert{}_\beta+n_\pert{}^\alpha \mu_\pert{}_\beta.
$

It is well known that
given the form of $\tidalpt$ in \eqref{tidal:tidalpt}, and thus of $g_\pert=g+\pert \tidalpt$,
the resulting $G^{(1)}$ takes a form so that the equations \eqref{eq:efes_first}
in both the perfect fluid and two-fluid cases for $\ell\geq 2$ yield \cite{Thorne1967,Char2018}

\begin{align}
&H_2{}_\ell=H_0{}_\ell,\label{tidal:H2} \\
&K_\ell{}' = H_0{}_\ell'+ \nu{}' H_0{}_\ell,\label{tidal:dK}\\
&r^2 \nu{}' H_0{}_\ell' = e^{\lambda} \left (\ell(\ell+1) -2 \right)K_\ell + \left(r (\lambda{}'+\nu{}') - \left(r \nu{}'\right)^2 -e^{\lambda} \ell(\ell+1) + 2 \right)H_0{}_\ell,\label{tidal:dH0} 
\end{align}
plus relations that involve the first order quantities relative
to the matter fluids.

We can now take differences of these equations at either side
of the matching hypersurfaces $\Sigma_{\ith}$ and thus obtain
relations between the jumps of the perturbation metric functions.
The task is then to combine the perturbed matching conditions we already
have with those relations.\\

\noindent
\emph{Perfect fluid:}
The combination of the set of equations \eqref{eq:matching_1}-\eqref{eq:matching_3}
with the differences of \eqref{tidal:H2}-\eqref{tidal:dH0},
using the relations in \eqref{eq:dlambda}-\eqref{eq:ddnu},
is equivalent to the set of equations \cite{reina-sanchis-vera-font2017}
\begin{align}
    &[M]_i=0,\quad[\nu]_i=0,\label{eq:matching_1_f}\\
  &[P]_i=0, \label{eq:matching_pressure_f}\\
     &[H_{0\ell}]_i=0,\quad[H_{2\ell}]_i=0,\quad[K_\ell]_i=0,
     \label{tidal:matching_tidal_PF_1}\\
    &[H'_{0\ell}]_i=[K'_\ell]_i=\frac{\varkappa e^{\lambda(R_i)}}{\nu'(R_i)}H_{0\ell}(R_i)[E]_i,\label{tidal:matching_tidal_PF_2}\\
    &[E]_i\left(\defor_{\ell i}+\frac{e^{\lambda(R_i)/2}}{\nu'(R_i)}H_{0\ell}(R_i)\right)=0,\label{tidal:matching_tidal_PF_3}
\end{align}
for $\ell\geq 2$.

Two comments are in order. First,
these matching conditions, for $\ell\geq 2$, appeared already
(without proof) in an analogous context in Appendix B in \cite{Price_Thorne_1969}. There, the matching
to all $\ell$ is included, although the ``continuity"
of the $\ell=0,1$ sector may need some partial gauge fixing \cite{ReinaVera2015, MRV2}.
Second, it must be stressed that
the background matching conditions \eqref{eq:matching_1_f} and
\eqref{eq:matching_pressure_f}, together with the ``continuity" conditions
\eqref{tidal:matching_tidal_PF_1}, plus the field equations \eqref{tidal:dK}
and \eqref{tidal:dH0} guarantee that the matching condition
\eqref{tidal:matching_tidal_PF_2} is satisfied.
This fact, which was also already observed in \cite{Price_Thorne_1969},
is the reason why the discontinuity of $[H_0']$ can be found
using the EFEs \emph{provided that the functions involved
are asummed to be continuous}.\\

\noindent
\emph{Superfluid:}
For the two-fluid model it is easy to see that it suffices to replace 
$E$ with $-\Lambda$
and $P$ with $\Psi$
in \eqref{eq:matching_SF_psi} and \eqref{eq:matching_SF_defor}.
As a result, the full set of perturbed matching conditions to first order in this case
is given by \eqref{eq:matching_1_f}, \eqref{tidal:matching_tidal_PF_1}
plus
\begin{align}
  &[\Psi]_i=0,\label{eq:matching_SF_psi_f}\\
  &[H'_{0\ell}]_i=[K'_\ell]_i=-\frac{\varkappa e^{\lambda(R_i)}}{\nu'(R_i)}H_{0\ell}(R_i)[\Lambda]_i,\label{tidal:matching_tidal_SF_2}\\
  &[\Lambda]_i\left(\defor_{\ell i}+\frac{e^{\lambda(R_i)/2}}{\nu'(R_i)}H_{0\ell}(R_i)\right)=0,\label{tidal:matching_tidal_SF_3}
\end{align}
for $\ell\geq2$.\\

\noindent
\emph{Superfluid and perfect fluid:}
Likewise, for an inner superfluid model matching to an outer perfect fluid,
the set is given by \eqref{eq:matching_1_f}, \eqref{tidal:matching_tidal_PF_1}
plus
\begin{align}
  &\Psi(R_i)=P(R_i),\label{eq:matching_SF_PF_1_f}\\
    &[H'_{0\ell}]_i=[K'_\ell]_i=-\frac{\varkappa e^{\lambda(R_i)}}{\nu'(R_i)}H_{0\ell}(R_i)\left(\Lambda(R_i)+E(R_i)\right),\label{tidal:matching_tidal_SF_PF_2}\\
    &\left(\Lambda(R_i)+E(R_i)\right)\left(\defor_{\ell i}+\frac{e^{\lambda(R_i)/2}}{\nu'(R_i)}H_{0\ell}(R_i)\right)=0,\label{tidal:matching_tidal_SF_PF_3}
\end{align}
for $\ell\geq2$.\\

Clearly, the same equations with a change of sign in $E$ and $\Lambda$
hold if the inner region is a perfect fluid and the outer region is the superfluid.

%----------------------------
\section{Tidal problem and Love number}
\label{sec:love_number}

The Love number at each harmonic $\ell$ is obtained from the function
$H_{0\ell}$ and its radial derivative $H'_{0\ell}$ evaluated
at the outermost boundary of the star in the vacuum exterior,
that we denote by $\manifold^-$ with relative quantities
also tagged with the minus sign.
In the present setting and notation, that is
$H^-_{0\ell}(\rstar)$ and $H'{}^-_{0\ell}(\rstar)$.
The solution $H_{0\ell}$
of the equations \eqref{tidal:dK}-\eqref{tidal:dH0} for a vacuum region are the associated Legendre functions $P_\ell^n$, $Q_\ell^n$ with $n = 2$
\begin{align}
    H_{0\ell}^-(r_-)=a_{\ell P}\hat{P}_{\ell}^2\left(\frac{r_-}{M}-1\right)+a_{\ell Q}\hat{Q}_{\ell}^2\left(\frac{r_-}{M}-1\right),\label{eq:tidal_ext}
\end{align}
where the Legendre functions have been normalized so that $\hat{P}_\ell^2(x)\simeq x^{\ell}$ and $\hat{Q}_\ell^2(x)\simeq 1/x^{\ell+1}$ when $x\rightarrow\infty$, and are given by
\begin{align*}
    \hat{P}_{\ell}^2(x):=\left(\frac{2^\ell}{\sqrt{\pi}}\frac{\Gamma(\ell+1/2)}{\Gamma(\ell-1)}\right)^{-1} P_{\ell}^2(x),\\
    \hat{Q}_{\ell}^2(x):=\left(\frac{\sqrt{\pi}}{2^{\ell+1}}\frac{\Gamma(\ell+3)}{\Gamma(\ell+3/2)}\right)^{-1} Q_{\ell}^2(x).
\end{align*}

The Love number is defined as
\begin{align}
    k_\ell:=\frac{1}{2}\left(\frac{M}{\rstar}\right)^{2\ell+1}a_\ell,\label{love:love}
\end{align}
where $M$ is the mass of the star $M=M(\rstar)$
(observe the mass function has no jumps, c.f. \eqref{eq:matching_1_f}), and
$a_\ell$ is the ratio between the two constants in \eqref{eq:tidal_ext}
\begin{align}
    a_\ell:=\frac{a_{\ell Q}}{a_{\ell P}}=-\frac{\partial_{r_-} \hat{P}_\ell^2-(y_\ell^-/\rstar)\hat{P}_\ell^2}{\partial_{r_-} \hat{Q}_\ell^2-(y_\ell^-/\rstar)\hat{Q}_\ell^2}\bigg\lvert_{r_-=\rstar},\label{love:a_l}
\end{align}
where $y_\ell:=rH_{0\ell}'/H_{0\ell}$. The leading order Love number, $k_2$, is many times replaced
(but often called also Love number) by the tidal deformability
$\lambda_2=a_2/3$.

Computationally, one must fix some boundary conditions at the origin
and integrate the EFE from $\manifold_{(1)}$ to $\manifold_{(N)}$,
respecting the matching conditions at each boundary, until reaching
the outermost boundary, $\Sigma_{(N)}$. Once the interior problem
has been solved (and thus $y_\ell^+(\rstar)$ is known), the value of
$y_\ell^-$ at the boundary, $y_\ell^-(\rstar)$,
has to be obtained from the matching of the two regions,
$$y_\ell^-(\rstar)=y_\ell^+(\rstar)-[y_\ell]_N$$ with
\begin{align}
    [y_\ell]_N=\left[\frac{r H_{0\ell}'}{H_{0\ell}}\right]_N
    =\frac{\rstar}{H_{0\ell}(\rstar)}\left[H'_{0\ell}\right]_N,\label{love:ysalto}
\end{align}
after using $[H_{0\ell}]_N=0$ in the last equality.
Then, using \eqref{love:a_l}
the tidal deformability, $\lambda_2$ (and the Love number $k_2$
using also \eqref{love:love}) is computed.\\

\noindent
\emph{Perfect fluid:}
For the perfect fluid case,
using equations \eqref{tidal:matching_tidal_PF_1} and \eqref{tidal:matching_tidal_PF_2}, together with the vacuum solutions of $\lambda(r)$ and $\nu(r)$,
\begin{align}
    e^{-\lambda(r)}=e^{\nu(r)}=1-\frac{2M}{r},
\end{align}
equation~\eqref{love:ysalto} reads
\begin{align}
    [y_\ell]_N=\varkappa\frac{\rstar^3}{2M}E(\rstar),
\end{align}
thus recovering \eqref{eq:matching_y} for $\ell=2$ after denoting $\rstar=\ro$.
Then,
\begin{align}
    a_\ell=-\frac{\partial_{r_+} \hat{P}_\ell^2-(y_\ell^+/\rstar)\hat{P}_\ell^2+(\varkappa \rstar^2 E(\rstar)/2M)\hat{P}_\ell^2}{\partial_{r_+} \hat{Q}_\ell^2-(y_\ell^+/\rstar)\hat{Q}_\ell^2+(\varkappa \rstar^2 E(\rstar)/2M)\hat{Q}_\ell^2}\bigg\lvert_{r_+=\rstar},
\end{align}
where we have recovered the $+$ index in $y_\ell^+$ because $y_\ell^+\neq y_\ell^-$ in general.\\

\noindent
\emph{Superfluid:}
For the superfluid case, the analogous procedure
shows that it suffices to replace $E$ by $-\Lambda$,
so that \cite{Aranguren:tidal}
\begin{align}\label{eq:jump_y}
    [y_\ell]_N=-\varkappa\frac{\rstar^3}{2 M}\Lambda(\rstar),
\end{align}
and
\begin{align}
    a_\ell&=-\frac{\partial_{r_+} \hat{P}_\ell^2-(y_\ell^+/\rstar)\hat{P}_\ell^2-(\varkappa \rstar^2\Lambda(\rstar)/2M)\hat{P}_\ell^2}{\partial_{r_+} \hat{Q}_\ell^2-(y_\ell^+/\rstar)\hat{Q}_\ell^2-(\varkappa \rstar^2\Lambda(\rstar)/2M)\hat{Q}_\ell^2}\bigg\lvert_{r_+=\rstar}.\label{eq:a_l}
\end{align}

\section{Conclusions}

The set of geometrical matching conditions reviewed in this article serves as a basis for those works that aim to study the tidal problem for non perfect fluid stellar configurations for which Eq.~(15) in \cite{Hinderer:2010} no longer applies as such. In this regard, although equations to first order for perfect fluid and superfluid interiors appear to be analogous (in the matching conditions it is just a change $E\rightarrow-\Lambda$ and $P\rightarrow\Psi$), at second order this similarity is completely lost; compare, for instance, the contribution of the mass at second order, $\delta M$, from Eq.~(103) in \cite{ReinaVera2015} with that of Eq.~(97) in \cite{Aranguren:2022}.

Even though we have restricted ourselves to the theory of General Relativity using the Einstein field equations at some point, due to the geometrical nature of \eqref{eq:matching_background1}-\eqref{tidal:s_dH0_dK_2}, which are a consequence of imposing that the Riemann tensor has no Dirac delta terms, these equations may still hold as a minimum set of matching conditions on alternative metric theories of gravity, as well as metric-affine theories
(except for one very particular case, see \cite{Casado_Turrion_2023}). In those alternative theories one would
expect to need more conditions, as it happens when dealing
with quadratic theories of gravity, see e.g. \cite{Reina_2016} and
references therein. Let us stress finally that the
geometrical procedure in Sec.~\ref{sec:geometrical}
is suited to be applied to any stellar exterior,
that is, regardless of whether the ``vacuum" exterior
of some given theory is or is not Schwarzschild, see e.g. \cite{APARICIORESCO2016147}.

\section*{Acknowledgments}
We thank the anonymous referee for useful suggestions. Work supported by the Basque Government (Grant No. IT1628-22) and Spanish Agencia Estatal de Investigación (Grant No. PID2021-123226NB-I00 funded by “ERDF A way of making Europe” and MCIN/AEI/10.13039/501100011033). E. A. is supported by the Basque Government Grant No. PRE\_2023\_2\_0199.

\bibliography{super_fluid}

\begin{thebibliography}{10}

\bibitem{Damour:2009vw}
T.~Damour and A.~Nagar, ``{Relativistic tidal properties of neutron stars},''
  {\em Phys. Rev. D}, vol.~80, p.~084035, 2009.

\bibitem{Hinderer:2010}
T.~Hinderer, B.~D. Lackey, R.~N. Lang, and J.~S. Read, ``Tidal deformability of
  neutron stars with realistic equations of state and their gravitational wave
  signatures in binary inspiral,'' {\em Phys. Rev. D}, vol.~81, p.~123016, Jun
  2010.

\bibitem{reina-sanchis-vera-font2017}
B.~Reina, N.~Sanchis-Gual, R.~Vera, and J.~A. Font, ``{Completion of the
  universal I-Love-Q relations in compact stars including the mass},'' {\em
  \mnras}, vol.~470, pp.~L54--L58, 2017.

\bibitem{Aranguren:tidal}
E.~{Aranguren}, J.~A. {Font}, N.~{Sanchis-Gual}, and R.~{Vera}, ``{Revisiting
  the I -Love-Q relations for superfluid neutron stars},'' {\em \prd},
  vol.~108, p.~104065, Nov. 2023.

\bibitem{Hinderer:2008}
T.~Hinderer, ``{Tidal Love numbers of neutron stars},'' {\em Astrophys. J.},
  vol.~677, pp.~1216--1220, 2008.

\bibitem{regge:1957}
T.~Regge and J.~A. Wheeler, ``Stability of a schwarzschild singularity,'' {\em
  Phys. Rev.}, vol.~108, pp.~1063--1069, Nov 1957.

\bibitem{Hartle1967}
J.~B. Hartle, ``Slowly rotating relativistic stars. i. equations of
  structure,'' {\em \apj}, vol.~150, pp.~1005--1029, 1967.

\bibitem{Hartle2}
J.~B. {Hartle} and K.~S. {Thorne}, ``{Slowly Rotating Relativistic Stars. II.
  Models for Neutron Stars and Supermassive Stars},'' {\em \apj}, vol.~153,
  p.~807, Sept. 1968.

\bibitem{ReinaVera2015}
B.~Reina and R.~Vera, ``Revisiting hartle's model using perturbed matching
  theory to second order: amending the change in mass,'' {\em \CQG}, vol.~32,
  p.~155008, jul 2015.

\bibitem{Marc_Filipe_Raul_2007}
M.~Mars, F.~C. Mena, and R.~Vera, ``Linear perturbations of matched spacetimes:
  the gauge problem and background symmetries,'' {\em \CQG}, vol.~24, p.~3673,
  jul 2007.

\bibitem{MRV1}
M.~{Mars}, B.~{Reina}, and R.~{Vera}, ``{Gauge fixing and regularity of axially
  symmetric and axistationary second order perturbations around spherical
  backgrounds},'' {\em Adv. Theo. Math. Phys.}, vol.~26, pp.~1873--1963, 2022.

\bibitem{MRV2}
M.~{Mars}, B.~{Reina}, and R.~{Vera}, ``{Existence and uniqueness of compact
  rotating configurations in GR in second order perturbation theory},'' {\em
  Adv. Theo. Math. Phys.}, vol.~26, pp.~2719--2840, 2022.

\bibitem{Mars:2005}
M.~{Mars}, ``{First- and second-order perturbations of hypersurfaces},'' {\em
  \CQG}, vol.~22, pp.~3325--3347, Aug. 2005.

\bibitem{Battye01}
R.~A. Battye and B.~Carter, ``Generic junction conditions in brane-world
  scenarios,'' {\em \PLB}, vol.~509, no.~3-4, p.~331, 2001.

\bibitem{Mukohyama00}
S.~Mukohyama, ``Gauge-invariant gravitational perturbations of maximally
  symmetric spacetimes,'' {\em \PRD}, vol.~62, p.~084015, oct 2000.

\bibitem{JMaria-Carsten}
J.~M. Mart\'{\i}n-Garc\'{\i}a and C.~Gundlach, ``Gauge-invariant and
  coordinate-independent perturbations of stellar collapse. ii. matching to the
  exterior,'' {\em Phys. Rev. D}, vol.~64, p.~024012, Jun 2001.

\bibitem{Vera2002}
R.~Vera, ``{Symmetry-preserving matchings},'' {\em \CQG}, vol.~19,
  pp.~5249--5264, oct 2002.

\bibitem{Aranguren:2022}
E.~Aranguren, J.~A. Font, N.~Sanchis-Gual, and R.~Vera, ``Revised formalism for
  slowly rotating superfluid neutron stars in general relativity,'' {\em Phys.
  Rev. D}, vol.~107, p.~044034, Feb 2023.

\bibitem{Reina:2015jia}
B.~Reina, ``{Slowly rotating homogeneous masses revisited},'' {\em Mon. Not.
  Roy. Astron. Soc.}, vol.~455, no.~4, pp.~4512--4517, 2016.

\bibitem{alberto_skyrme}
C.~Adam, A.~G. Mart\'{\i}n-Caro, M.~Huidobro, R.~V\'azquez, and
  A.~Wereszczynski, ``Quasiuniversal relations for generalized skyrme stars,''
  {\em Phys. Rev. D}, vol.~103, p.~023022, Jan 2021.

\bibitem{Langlois:1998}
D.~{Langlois}, D.~M. {Sedrakian}, and B.~{Carter}, ``{Differential rotation of
  relativistic superfluid in neutron stars},'' {\em \mnras}, vol.~297,
  pp.~1189--1201, July 1998.

\bibitem{Comer1999}
G.~L. Comer, D.~Langlois, and L.~M. Lin, ``{Quasinormal modes of general
  relativistic superfluid neutron stars},'' {\em Phys. Rev. D}, vol.~60,
  p.~104025, 1999.

\bibitem{datta_char:2020}
S.~{Datta} and P.~{Char}, ``{Effect of superfluid matter of a neutron star on
  the tidal deformability},'' {\em \prd}, vol.~101, p.~064016, Mar. 2020.

\bibitem{Thorne1967}
K.~S. Thorne and A.~Campolattaro, ``Non-radial pulsation of
  general-relativistic stellar models. i. analytic analysis for l $\geq$ 2,''
  {\em The Astrophysical Journal}, vol.~149, p.~591, sep 1967.

\bibitem{Char2018}
P.~Char and S.~Datta, ``{Relativistic tidal properties of superfluid neutron
  stars},'' {\em Phys. Rev. D}, vol.~98, no.~8, pp.~1--18, 2018.

\bibitem{Price_Thorne_1969}
R.~{Price} and K.~S. {Thorne}, ``{Non-Radial Pulsation of General-Relativistic
  Stellar Models. II. Properties of the Gravitational Waves},'' {\em \apj},
  vol.~155, p.~163, Jan. 1969.

\bibitem{Casado_Turrion_2023}
A.~Casado-Turrión, Álvaro {de la Cruz-Dombriz}, A.~{Jiménez Cano}, and F.~J.
  {Maldonado Torralba}, ``Junction conditions in bi-scalar poincaré gauge
  gravity,'' {\em Journal of Cosmology and Astroparticle Physics}, vol.~2023,
  p.~023, jul 2023.

\bibitem{Reina_2016}
B.~Reina, J.~M.~M. Senovilla, and R.~Vera, ``Junction conditions in quadratic
  gravity: thin shells and double layers,'' {\em \CQG}, vol.~33, p.~105008, apr
  2016.

\bibitem{APARICIORESCO2016147}
M.~{Aparicio Resco}, Álvaro {de la Cruz-Dombriz}, F.~J. {Llanes Estrada}, and
  V.~{Zapatero Castrillo}, ``On neutron stars in f(r) theories: Small radii,
  large masses and large energy emitted in a merger,'' {\em Physics of the Dark
  Universe}, vol.~13, pp.~147--161, 2016.

\end{thebibliography}
\bibliographystyle{ieeetr}

\end{document}